\RequirePackage{fix-cm}
\documentclass[smallextended]{svjour3}       
\smartqed  
\usepackage{graphicx}
 \journalname{my journal}
\newcommand{\aap}{{Astron. Astrophys.}}

\newcommand{\solphys}{{Solar Phys.}}
\begin{document}

\title{Solar activity in the past and the chaotic behaviour of the dynamo
}


\author{Rainer Arlt         \and
        Nigel Weiss 
}


\institute{R. Arlt \at
              Leibniz Institute for Astrophysics Potsdam\\
              Tel.: +49-331-7499-354\\
              Fax: +49-331-7499-526\\
              \email{rarlt@aip.de}           
           \and
           N. Weiss \at
              Department of Applied Mathematics and Theoretical Physics \\
              University of Cambridge, UK \\
              \email{now1@cam.ac.uk}           
}

\date{Received: date / Accepted: date}

\maketitle

\begin{abstract}
The record of solar activity is reviewed here with emphasis on
peculiarities. Since sunspot positions tell us a lot more about 
the solar dynamo than the various global sunspot numbers, we 
first focus on the records of telescopic observations of 
sunspots leading to positional information.
Then we turn to the proxy record from cosmogenic isotope abundances, which shows
recurrent grand minima over the last 9500 years. The apparent distinction between
episodes of strong modulation, and intervening episodes with milder modulation and
weaker overall activity, hints at the solar dynamo following a variety
of solutions, with different symmetries, over the course of millennia.

\end{abstract}

\section{Introduction}
\label{intro}
Telescopic observations of sunspots have revealed both the 11-year Schwabe cycle and
the interruption of activity associated with the Maunder Minimum in the 17th century.
New analyses of early records (including those of Schwabe) confirm that the pattern
associated with the butterfly diagram has been present for the past 300 years.  There
is also evidence of differential rotation, with suggestions of anomalous behaviour
during the Maunder Minimum.

The record of solar activity has been extended back for almost 10\,000 years by
measuring the abundances of the cosmogenic isotopes $^{10}$Be and $^{14}$C in ice
cores and tree rings.  This record reveals many grand minima, with a characteristic
spacing of around 200 years (the de Vries cycle) but the appearance of these grand
minima itself varies with a characteristic timescale of over 2000 years.  We interpret
the grand minima and maxima as resulting from deterministic modulation of the nonlinear 
solar dynamo, oscillating chaotically with the mean Hale period of around 22 years. 
The present peculiar evolution of the solar cycle may turn out to be an enlightening
period in this respect as well, since it will be observed by various methods including
helioseismology.

Although the Sun's magnetic field is now predominantly dipolar, simple
nonlinear models show that symmetry can flip to give quadrupolar or even mixed-mode
behaviour.  We suggest that such flipping explains the long-term, multimillennial 
variability of the activity record.

\section{The sunspot record}
\label{sec:1}

The sunspot number goes back to Wolf (1859) who defined an index
of solar activity based on the number of sunspot groups and the
total number of individual spots on the observable 
hemisphere of the Sun. The time series starts in 1749 with the 
observations by Johann Staudacher (Nuremberg) and has been continued 
in terms of the International Sunspot Number until the present day.
An alternative index was defined by Hoyt and Schatten (1998) who
only counted the group numbers -- an index that is more robust
against variable capabilities of seeing small spots and allowed
the time series to go back to the first days of telescopic observations
of the Sun in 1610.

The time-series are very often used as a proxy for some sort
of magnetic field strength in the interior of the Sun and are
compared with dynamo models. These global indices, however, cannot 
give any insight into the topology of the magnetic fields that
presumably generate the spots on the surface. One may imagine that
the knowledge of the heliographic positions of the spots can be
used to infer the equatorial symmetry, the latitudinal distribution,
and  the rotational symmetry of the underlying magnetic fields as 
well as the lifetime of magnetic structures on the solar surface. 

Sunspot positions are now being collected in a database using
results from a USAF network of observing stations, following on
the photoheliographic programme conducted by the Royal Greenwich
Observatory (RGO) which effectively also was a network of stations
around the world. The RGO/USAF set started in 1874 and stores only
the average position of sunspot groups together with their total
area. In parallel, several other programmes were collecting sunspot
positions of various amounts, or are still doing so.

\begin{figure}
\includegraphics[width=\textwidth]{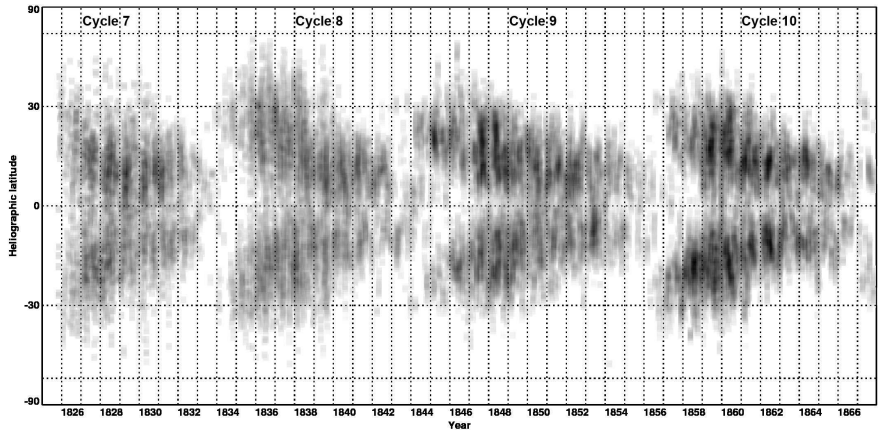}
\caption{Butterfly diagram obtained from the sunspot positions
derived from the drawings of Samuel Heinrich Schwabe in 1825--1867.
(After Arlt et al. 2013.)}
\label{butterfly_schwabe}
\end{figure}

Before that, a large set of sunspot data is available from Friedrich Wilhelm 
Gustav Sp\"orer who observed from 1861 to 1894 from the towns of Anklam and
Potsdam, Germany. He confirmed that the lull of reports on 
sunspots in the second half of the 17th century represented a real low
in solar activity for decades. His work was recognized later by
Maunder who was eventually credited with that discovery, whence 
the name ``Maunder minimum''. Sp\"orer's observations and 
measurements were published in a series of papers (Sp\"orer 1874, 
1878, 1880, 1886, 1894), but his original sunspot drawings -- 
if they existed -- are lost. 

Before Sp\"orer, Richard Carrington had already made the discovery
that the Sun is not rotating uniformly, but faster at the equator
than at higher latitudes (Carrington 1859). However, his observational data
only cover a rather short period, from November~1853 to March~1861
(Zolotova et al. 2010, Lepshokov et al. 2012).

A great extension of the butterfly diagram into the past comes 
from the observations by Samuel Heinrich Schwabe, who drew sunspots
in a solar disk each day he saw at least a glimpse of the Sun from
Nov~5, 1825, to Dec~31, 1867. He actually observed until Dec~15,
1868, but his last observing book was lost. Schwabe was also
the first to publish a paper suggesting that the sunspot number varies
periodic (Schwabe 1844). The positions and sizes of all sunspots
seen by Schwabe were measured by Arlt et al. (2013). The
resulting butterfly diagram is shown in Fig.~\ref{butterfly_schwabe}.

The time around 1800 is poorly covered by observations, the
longest record being that preserved by Honor\'e Flaugergues, who
reported useful sunspot observations in 1788--1830. These, however,
are yet to be analysed. They consist mostly of timings at a
transit instrument. Flaugergues gave transit times of the solar
limb and spots at a vertical and an oblique wire, which will
allow us to determine the latitudes of spots.

\begin{figure}
\includegraphics[width=\textwidth]{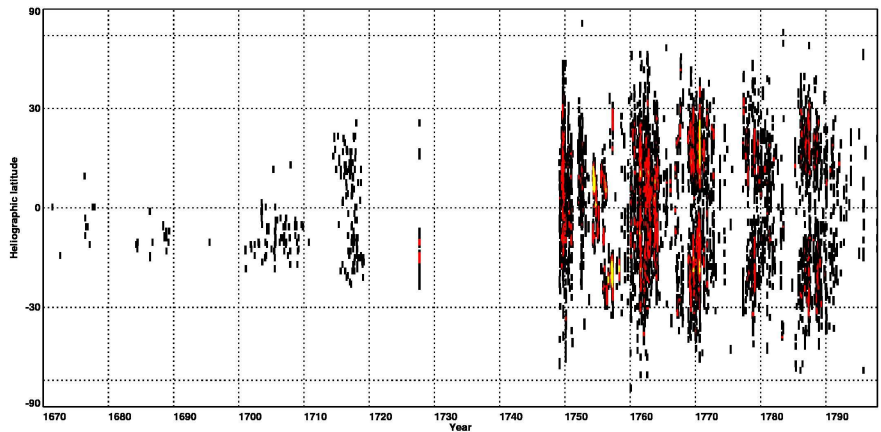}
\caption{Butterfly diagram of the period around and after
the Maunder minimum with sunspot positions from
various sources. The period until 1719 shows the positions
derived by Ribes and Nesme-Ribes (1993) which were digitized
by Vaquero et al. (http://haso.unex.es/); the year 1727 shows
two additional observations found at Paris observatory by
the authors. The period of 1749-1799 contains observations
by Staudacher (Arlt 2009a), Zucconi (Cristo et al. 2011),
and Hamilton (Arlt 2009b). Higher contrast is used than in
Fig.~\ref{butterfly_schwabe} because of the fewer spots available.}
\label{butterfly_maunder}
\end{figure}

A very interesting record of observations is stored in the library
of the Leibniz Institute for Astrophysics Potsdam. About 1000 drawings
of the Sun made by Johann Caspar Staudacher cover the period of 1749--1799.
The drawings are not accompanied by much verbal information about the
telescope or the observing method. There are no indications of the
orientations of the solar disks. Fortunately, detailed drawings of
partial solar eclipses showing the path and direction of the Moon
clearly show that the Sun was projected on a screen behind the 
telescope, i.e. all images are mirrored. The resulting butterfly 
diagram is shown in Fig.~\ref{butterfly_maunder} (Arlt 2009a).
It is remarkable that the first two cycles observed by Staudacher 
do not show a clear butterfly shape. Since the observer recorded
the sunspots with the projection method, they are certainly not
plotted `at random' into the disks, though the uncertainty of the
orientations holds true for the entire data set.

The observations were complemented by the very accurate drawings of 
Ludovico Zucconi in 1754--1760 (Zucconi 1760) which were analyzed 
by Cristo et al. (2011). They fill in the gaps of the observations 
by Staudacher around the minimum near 1755  and may help understand 
the unusual time--latitude distribution  of Staudacher's
spots. At the other end of Staudacher's observing period, additional
positions of a small number of sunspots were derived from the records
of Hamilton and Gimingham at Armagh Observatory in 1795--1797 (Arlt 2009b).

Further back in time, we find the analysis by Ribes and Nesme-Ribes (1993)
who measured the positions of sunspots seen by a variety of astronomers at 
the Observatoire de Paris, during and after the Maunder minimum, resulting
in data for 1672--1719 (Fig.~\ref{butterfly_maunder}). There is a striking absence of sunspots until
about 1714 with only the southern hemisphere being populated, by roughly
80~spots (plus 4~spots in the northern hemisphere and 6~spots right
on the equator). This result again shows that the solar cycle does
not necessarily show a butterfly shape for the time--latitude
distribution of spots. The actual activity cycle may have persisted
during the Maunder minimum as seen in the cosmic ray record (see Sect.~4)
at high time resolution (Beer et al. 1998; Berggren et al. 2009).

A fair number of observers recorded sunspots in the period before
the Maunder minimum, starting with the first telescopic
observations by Galileo Galilei and Thomas Harriot in 1610, followed
by Christoph Scheiner and Johannes Fabricius, who
first published the telescopic sunspot observations, in a printed
pamphlet (Fabricius 1611). There is no compilation of sunspot
positions for all the available sources yet, but visual inspection
of the images indicates normal spot distributions before the Maunder 
minimum.

\section{Results from the sunspot record}
\label{sec:2}

Sunspots show the latitudinal differential rotation of the Sun.
This has first been derived quantitatively by Carrington (1859)
and Peters (1859). The rotation of newly emerged sunspot groups
is, by the way, different from the rotation of the bulk gas at
the surface at the same latitude (Tuominen 1962, Pulkkinen and
Tuominen 1998).

Historical sunspot observations may actually allow measurements of
the differential rotation. A recent attempt by Arlt and Fr\"ohlich
(2012) employed Bayesian inference on the observations by Staudacher
to obtain positions, orientation angles and differential rotation 
parameter at the same time and delivered a latitudinal shear 
compatible with that of today. There is a slight but insignificant
hint that the differential rotation was stronger in the first
third of Staudacher's observations than during the remaining period.
This coincides with the period of non-butterfly shaped distribution 
of spot latitudes over time which is an automatic side product of 
the analysis.

The spot positions derived by Ribes and Nesme-Ribes (1993) also
indicate a stronger differential rotation at the end of the
Maunder minimum, along with an unusual spot distribution
as well. There is actually no fundamental reason why the 
solar magnetic field should always adopt a chief\/ly dipolar
structure (antisymmetric with respect to the equator). Quadrupolar
modes or mixed symmetries are certainly possible and may lead to 
butterfly diagrams deviating from the one we know today.

\begin{figure}
\begin{center}
\includegraphics[width=0.7\textwidth]{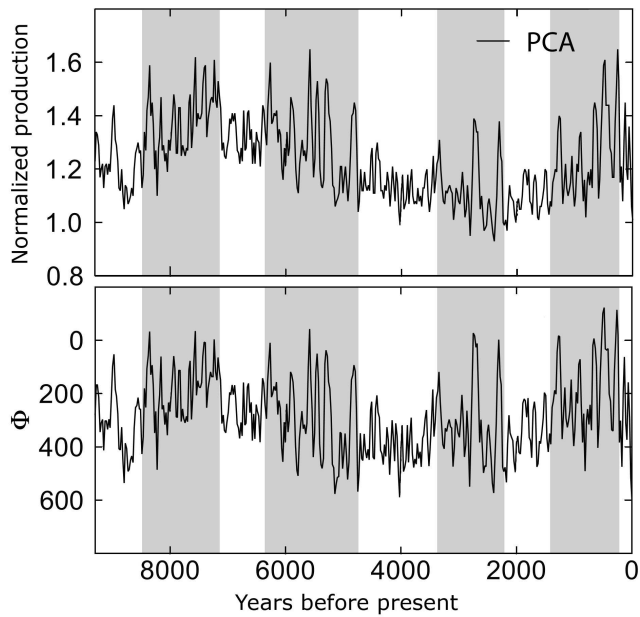}
\caption{The 9400-year record of solar magnetic activity derived from cosmogenic
isotopes for the period from 9400~BP to the present, where `present' means 1950.
Upper panel: the principal components record of the production rate, based on the
INTCAL09 record for $^{14}$C and GRIP and EDML abundances for $^{10}$Be. 
Lower panel: the modulation
function $\Phi$, in MeV, after correction in an attempt to eliminate the effects of
variations in the dipole moment of the geomagnetic field. The shaded strips denote the
intervals with vigorous modulation of solar magnetic activity, associated with the
presence of prominent grand maxima and grand minima.
(After McCracken et al. 2013.)}
\end{center}
\label{fig:McC1}
\end{figure}

Another issue of the solar cycle and the solar dynamo is the 
coupling between the hemispheres. Zolotova et al. (2010) 
studied the temporal variation of the phase lag between the
cycles separated into hemispheres. The phase difference varies
and changes sign approximately every 35--40~years, giving a
full period for the phase lag change of 70--80~years.

\section{The cosmic-ray record}

Galactic cosmic rays are deflected by magnetic fields in the heliosphere and so
the solar cycle modulates the flux of galactic cosmic rays into the 
Earth's atmosphere. The detection of the near-Earth cosmic
rays by decay products in the atmosphere (mostly neutrons)
delivered a 60-year record which shows a very good
anti-correlation with the sunspot record over the last five 
cycles -- see Usoskin (2013) for a review. Cosmic rays also lead to the production of
isotopes such as $^{14}$C and $^{10}$Be, which are absorbed into tree-rings or into
polar icecaps, where their abundances can be measured with great precision.
A 1000-year time-series of 
the isotope productions of $^{10}$Be and $^{14}$C (Muscheler et al. 2007) shows
good agreement with the sunspot record, although some significant differences remain.
More recently, the $^{14}$C data have been combined with $^{10}$Be measurements from
Greenland and
Antarctic ice cores to produce a much longer, composite time-series of cosmic
radiation with a duration of 9400~years. A Principal Components Analysis has then
been used to filter out most of the climatic effects and to reveal the presence of
recurrent grand maxima and grand minima
(Steinhilber et al. 2012; Abreu et al. 2013: McCracken et al. 2013).

The resulting time series can then be converted into a record of the modulation 
function $\Phi$, which can be very roughly interpreted as the mean loss of momentum-to-charge 
ratio of a cosmic ray particle in the heliosphere (Usoskin 2013).
Fig.~\ref{fig:McC1} displays the principal components rate of production of cosmogenic isotopes 
from 9400~BP to 1950~CE, together with the derived variation of the modulation
function $\Phi$, which has been corrected to take account of changes in the Earth's
magnetic field (Knudsen et al. 2008).

\begin{figure}
\begin{center}
\includegraphics[width=0.7\textwidth]{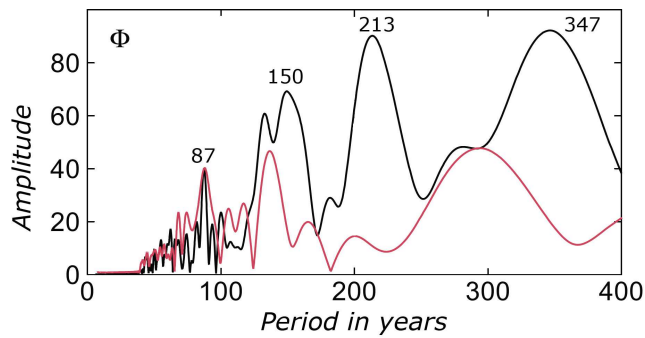}
\caption{Comparison of Fourier amplitude spectra for intervals with and without
grand minima.
Shown in black is the spectrum for $\Phi$ over the interval from
6300 to 4300~BP (containing four prominent grand minima).  The spectrum in red is for
the interval from 4700 to 3500, which contained no grand minima (as can be seen in
Fig.~\ref{fig:McC1}).  Apart from the Gleissberg peak at 87~yr and a possible
coincidence around 135~yr, the two spectra have little in common. 
(After McCracken et al. 2013.)}
\end{center}
\label{fig:McC6}
\end{figure}

The variations in the shaded regions of Fig.~\ref{fig:McC1} are all 
similar to those in the most recent millennium, which includes the Maunder, Sp\"orer,
Wolf and Oort Grand Minima.  Between these regions there are intervals, from
7100 to 6300~BP and from 4700 to 3500~BP, and again from 2200 to 1700~BP 
during which there are no grand minima and variations in isotope
production are relatively low.  This distinction becomes even more apparent if we
compare the Fourier amplitude spectra for intervals with and without grand extrema,
as displayed in Fig.~\ref{fig:McC6}.  The only clear coincidence is for the
Gleissberg cycle, with a period of 87~yr. The interval with strong modulation shows the
familiar de Vries period of 208~yr and other peaks at 150 and 350~yr, while the
2300~yr Hallstatt period shows up in the full record.  The intervals on either side,
with only weak modulation, have a broad peak around 300~yr and a sharper peak at
140~yr but the spectra are generally flatter.  All this confirms the immediate
impression that the behaviours in the shaded and unshaded regions of
Fig.~\ref{fig:McC1} are qualitatively different.    
The Hallstatt period then represents the characteristic timescale for transitions from
one regime to the other.

\section{Chaotic modulation and symmetry-breaking}

The Sun's cyclic activity is governed by macroscopic physics and is therefore
deterministic.  It is not practicable, however, to follow the emergence of every flux tube 
to form the sunspots that are shown in Fig.~\ref{butterfly_schwabe} and so we have to focus on 
averaged behaviour, which is deterministic but subject to stochastic disturbances.  Then it is
apparent that activity cycles must be regarded as manifestations of a chaotic
oscillator, with sensitive dependence on initial conditions (e.g. Zeldovich et al.
1983; Spiegel 2009).  The evolution of such a system is represented by a
trajectory in phase space; provided the stochastic perturbations are not too large, the
disturbed trajectories are always {\em shadowed\/} by nearby trajectories of the
undisturbed chaotic system (Ott 1993).  Turning to the observational records described
above, we should therefore expect the chaotic system to generate modulation
corresponding to grand minima and grand maxima, whose origin can be understood by
reference to the mathematical structure of the problem (Tobias and Weiss 2007). 
Simple models do indeed reproduce similar behaviour, which is associated with the
appearance of multiply periodic (``quasiperiodic'' to mathematicians) solutions and
global bifurcations that lead to chaos.

\begin{figure}
\begin{center}
\includegraphics[scale=0.5]{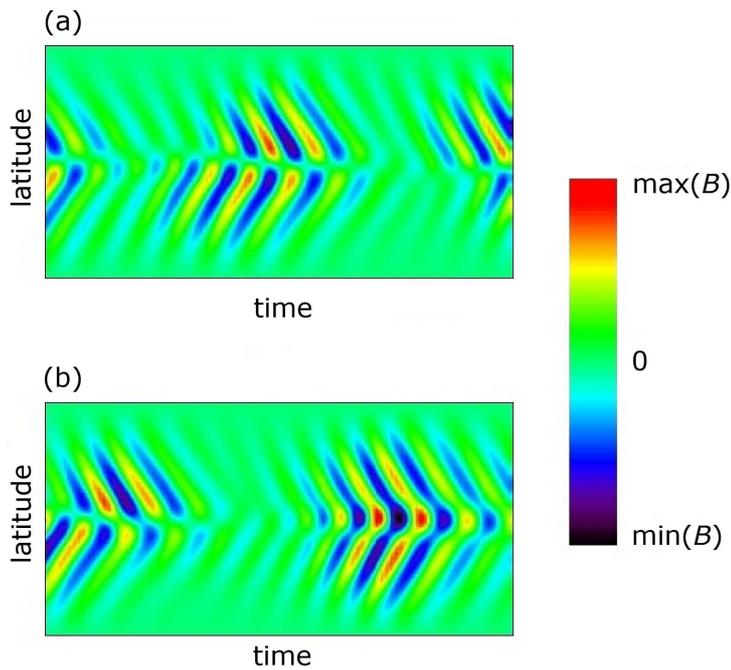}
\caption{Modulation and symmetry changes in a Cartesian mean-field dynamo
model governed by partial differential equations, showing the toroidal field
as a function of latitude and time.  (a) Active fields with dipole symmetry,
exhibiting grand minima associated with loss of symmetry and hemispheric
patterns.  (b) A transition from dipole symmetry to quadrupole symmetry
during a grand minimum (for the same parameter values).
(After Beer et al. 1998.)}
\end{center}
\label{fig:flippingpdes}
\end{figure}

\begin{figure}
\begin{center}
\includegraphics[scale=0.5]{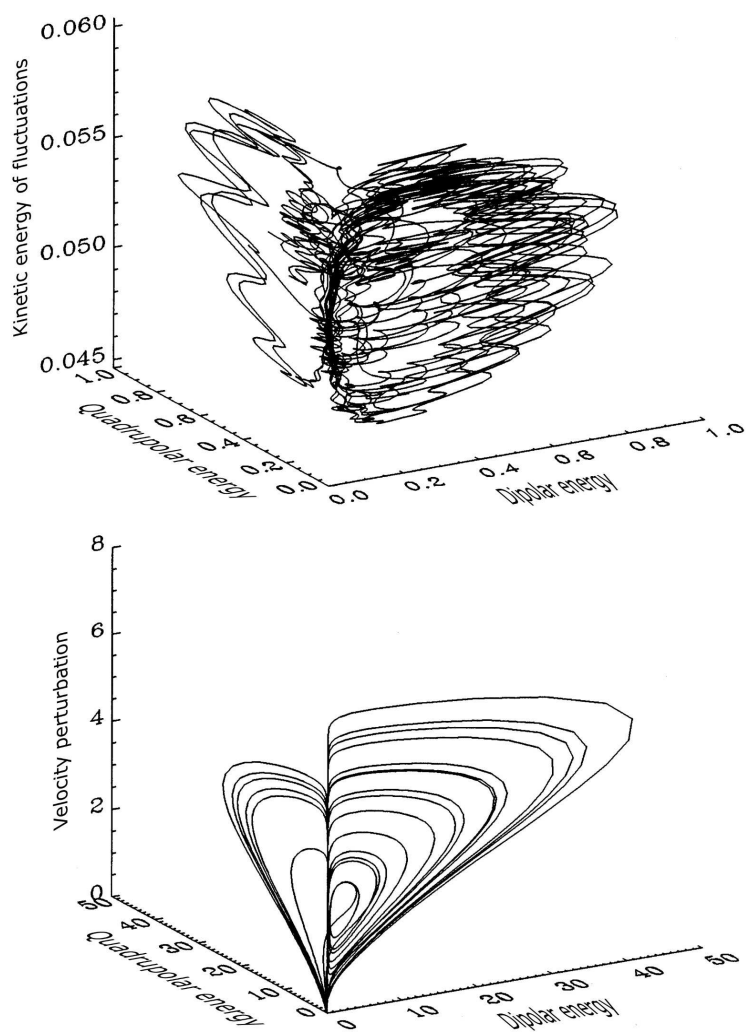}
\caption{Phase portraits illustrating flipping between dipole and quadrupole
polarities for simple model systems. Upper panel: a trajectory for the PDEs,
corresponding to Fig.~\ref{fig:flippingpdes} above, projected onto the 3-dimensional space
spanned by the dipole energy, the quadrupole energy and the perturbed
kinetic energy. Lower panel: the same but for the ODEs, with the perturbed
velocity as the ordinate.  In each case the symmetry flips occasionally at
deep grand minima.
(After Knobloch et al. 1998.)}
\end{center}
\label{fig:flippingports}
\end{figure}

In the simplest illustrative model, cyclic dynamo action sets in at an oscillatory (Hopf)
bifurcation that leads to periodic behaviour, with trajectories that are attracted to a
limit cycle in the phase space; this is followed by a second Hopf bifurcation that leads
to doubly periodic solutions that lie on a 2-torus in the 3-dimensional phase space;
after a series of period-doubling bifurcations (associated with a heteroclinic
bifurcation) behaviour becomes chaotic (Tobias, Weiss and Kirk 1995), though the chaotic
modulation still retains a memory of its original periodicity (Ott 1993).  Since this
bifurcation sequence was originally demonstrated for normal form equations (governing a
saddle-node/Hopf bifurcation) it is generic and therefore expected to be robust.
Indeed, the same pattern appears in simple dynamo models governed by partial
differential equations (Tobias 1996) and in mean-field dynamos (K\"uker et al. 1999; Pipin 1999; Bushby
2006).  It follows that grand minima and grand maxima should be interpreted as
{\em deterministic\/} effects, associated with chaotic modulation, and not as
products of large-scale stochastic disturbances.

All dynamo models allow two families of solutions that bifurcate from the trivial,
field-free state: these families have either dipole symmetry (with toroidal fields that
are antisymmetric about the equator) or quadrupole symmetry (with symmetric toroidal
fields).  Mixed modes can only appear as a result of symmetry-breaking bifurcations in
the nonlinear domain, which may lead to a complicated web of stable and unstable
solution branches (Jennings and Weiss 1992).  
After the Maunder minimum, the solar magnetic field appears to have gained dipolar 
symmetry by the second half of the 18th century, through a period of mixed symmetry 
(with nearly all spots in one hemisphere only) and a period poorly covered by observations.

Similar properties are exhibited by an idealized mean-field dynamo model, governed by
partial differential equations (Beer, Tobias and Weiss 1998), which also allows
transitions between dipolar and quadrupolar symmetries during grand minima.  This
behaviour is shown in Fig.~\ref{fig:flippingpdes}.  These properties are also demonstrated by even simpler
models, governed by low-order systems of ordinary differential equations (Knobloch et al.
1998).  Phase portraits for both PDEs and ODEs are displayed in Fig.~\ref{fig:flippingports}.
The former show both cyclic variations (predominantly horizontal) and large-amplitude
modulation, as well as occasional changes of symmetry.  With the ODEs it is possible to
filter out the cyclic variability, leaving only the modulation with flips of symmetry
near the origin, at very deep grand minima.  Note the reduced amplitude of modulation in
the quadrupole regime as compared with dipole fields.  By changing the parameters in the
model systems it is possible to find mixed-mode cycles too; they are likewise modulated,
and different symmetries may coexist without the possibility of flipping.

These results for highly simplified models reveal generic properties that would be shared
by solutions of the much more complicated equations that describe a real stellar dynamo.
What they show is not only that grand maxima and grand minima are associated with
deterministic modulation of cyclic activity but also that symmetry changes may provide a
natural explanation for the changes in behaviour that were reported by McCracken et al.
(2013) and summarized in Sections~2--4 above.  A more detailed discussion will appear
elsewhere (Tobias and Weiss, in preparation).

\begin{acknowledgements}
RA is grateful to the libraries of Leibniz Institute for Astrophysics Potsdam, the Royal
Astronomical Society, and the Observatoire de Paris for permitting the digitization of 
observing records. NW gratefully acknowledges many discussions with J\"urg Beer and 
Steven Tobias over the past two decades.
\end{acknowledgements}
\bibliographystyle{aps-nameyear}      


\end{document}